\documentclass[onecolumn]{emulateapj}
\usepackage{amsmath}
\usepackage{times}
\allowdisplaybreaks[1]

\def\eqq#1{Equation~\ref{#1}}
\newcommand\etal{{\it et al.\/}}
\newcommand\eg{{\it e.g.\/}}
\newcommand{\bfg}{\mbox{\boldmath $\gamma$}}
\newcommand{\bft}{\mbox{\boldmath $\theta$}}
\newcommand{\bfe}{\mbox{\bf e}}
\newcommand{\RR}{\mbox{$\cal R$}}

\newcommand{\Omb}{\mbox{$\Omega_{\rm b}$}}
\newcommand{\Omc}{\mbox{$\Omega_{\rm c}$}}
\newcommand{\Omm}{\mbox{$\Omega_{\rm m}$}}
\newcommand{\Omk}{\mbox{$\Omega_{\rm k}$}}
\newcommand{\Omde}{\mbox{$\Omega_{\rm de}$}}
\newcommand{\As}{\mbox{$A_{\rm s}$}}
\def\simlt{\lesssim}

\begin{document}

\title{Dark Energy Constraints from the CTIO Lensing Survey}

\author{Mike Jarvis, Bhuvnesh Jain, Gary Bernstein, Derek Dolney}
\affil{Dept. of Physics and Astronomy, University of Pennsylvania,
Philadelphia, PA 19104}
\email{mjarvis, bjain, garyb, dolney@physics.upenn.edu}

\begin{abstract}

We perform a cosmological parameter analysis of the 75 square
degree CTIO lensing survey in conjunction with CMB and Type Ia supernovae 
data.  For $\Lambda$CDM cosmologies,
we find that the amplitude of the power spectrum at low redshift is given by 
$\sigma_8 = 0.81^{+0.15}_{-0.10}\ (95\%\ {\rm c.l.})$, where the
error bar includes both statistical and systematic errors.
The total of all systematic errors is smaller than the statistical errors,
but they do make up a significant fraction of the error budget. 
We find that weak lensing improves the constraints on dark energy as well. 
The (constant) dark energy equation of state parameter, $w$, is
measured to be $-0.89^{+0.16}_{-0.21}\ (95\%\ {\rm c.l.})$. 
Marginalizing over a constant $w$ slightly changes the 
estimate of $\sigma_8$ to $0.79^{+0.17}_{-0.14}\ (95\%\ {\rm c.l.})$.
We also investigate variable $w$ cosmologies, but find that the constraints
weaken considerably; the next generation surveys are
needed to obtain meaningful constraints on the possible time evolution
of dark energy. 

\end{abstract}


\section{Introduction}

Observations of 
weak gravitational lensing, the coherent distortion in the images
of distant galaxies, have advanced rapidly in the past four years. 
The first detections of weak lensing in blank fields were reported
only a few years ago \citep{Wi00,KWL00,vW00,Ba00,Rh00}. 
More recent lensing measurements 
\citep{Ho02,Re02,Ba03,Br03,Ha03,Ja03,vW04,He05} 
have used larger and/or deeper surveys to reduce the statistical
errors, which scale as $N^{-1/2}$ where $N$ is the number of galaxies
measured. Better techniques for reducing systematic errors have also 
been developed, resulting in interesting cosmological constraints
from lensing surveys.  

Shear correlations measured by lensing surveys determine the 
projected power spectrum of matter fluctuations in the Universe. 
These fluctuations are believed to have grown due to gravitational
instability from the early universe to the present. The growth of
fluctations from last scattering, $z=1100$, to the present is 
sensitive to the densities of dark energy and matter via the Hubble
expansion rate. Further, the measured lensing signal depends on 
angular diameter distances to the source galaxies. Thus weak lensing
observables probe the dark energy density through both the growth function
and the geometric distance factors. Present lensing measurements 
are sensitive to redshifts $0 \simlt z\simlt 1$. 

As the size of weak lensing surveys increases and the statistical errors
keep going down, it becomes more important to similarly reduce 
the systematic errors.  
There are a few different systematic errors which can contaminate
a weak lensing signal at the level of typical cosmic shear measurements.
The largest of these has typically been the corrections of the anisotropic
point-spread-function (PSF).  We have recently developed a principal 
component analysis approach to interpolating the PSF between the stars
in the image using information from multiple exposures \citep{Ja05}. 
The PSF pattern is found to be a function of only a few underlying 
variables.  Therefore, we are able to improve the fits of this pattern
by using stars from many exposures. We have applied it to the 
CTIO survey data presented in \citet{Ja03} and have shown
that the level of systematic error is well below the statistical
errors.
Indeed the measured $B$-mode in shear correlations is consistent
with zero on all scales measured. 
In this paper we present a new parameter analysis of the CTIO 
lensing survey. With the reduced systematic error, we are able to extract
information from the measured signal over nearly two decades in length
scale. 

Type Ia supernovae observations led to the discovery of the
accelerated expansion of the universe (Riess \etal 1998; Perlmutter
\etal 1999). By combining information from
the CMB at $z=1100$, large-scale structure and SNIa, interesting
constraints on dark energy have been obtained
\citep{Sp03,bridle,We03,alam,saini,Te04,Wa04,Se05,simon,rapetti,Jas05}. 
Whether the dark energy
density is constant or evolving with cosmic time is one of the most
interesting observational questions. It is often expressed using the 
parameterization 
\begin{equation}
w(a) = w_0 + w_a(1-a), 
\label{eqn:wa}
\end{equation}
where $w$ is the dark energy equation of state parameter $w=p/\rho$
and $a=1/(1+z)$ is the expansion scale factor \citep{Ch01,Li03}. 
The cosmological constant corresponds to $w_0=-1, w_a=0$. 
A non-zero value of $w_a$ corresponds to a time-dependent equation of
state. 
We investigate three priors for $w$: $\Lambda$CDM ($w_0=-1, w_a=0$), 
constant $w$ ($w_a=0$) with $-3 < w < 0$, and variable $w$ with
$-8 < w < 8$ and $-8 < w_a < 8$.

In \S\ref{datasection} we briefly review our CTIO survey data and
shape measurement technique, largely deferring to our previous papers 
\citep{Ja03,Ja05} for more details.  
In \S\ref{statssection}, we present the results from the new
reduction.  We use these results to constrain cosmological parameters
in \S\ref{analysissection}, and conclude in \S\ref{discussionsection}.

\section{Data}
\label{datasection}

Our CTIO survey data was originally described in detail in \citet{Ja03},
and we refer the reader to that paper for most of the details about the
data and the analysis.  Here, we present a brief summary of the data set, 
and point out two significant changes in the analysis: our PSF interpolation
and the dilution correction.

The data were taken at Cerro Tololo Interamerican Observatory (CTIO) in Chile
from December, 1996 to July, 2000.  We observed 12 fields, well separated
on the sky, 
in low extinction but otherwise random locations.
Each field is approximately 2.5 degrees on a side, giving us a total of 
75 square degrees.  The 50\% completeness level occurs near $R = 23.5$
for each field, although it varies somewhat between the fields. 
We impose limits of $19 < R < 23$, which gives about 2 million 
galaxies to use for our lensing statistics.

The shape measurements of the galaxies follow the techniques of 
\citet{BJ02}.  The galaxy shapes are measured using an elliptical Gaussian 
weight function which is matched to have the same ellipticity as the galaxy.
(That is, we use a circular Gaussian in the sheared coordinate system in 
which the galaxy is round.)  The observed ellipticity is then calculated as:
\begin{equation}
\bfe = e_1 + i e_2 = 
\frac{\int I(x,y) W(x,y) (x+iy)^2 dxdy}{\int I(x,y) W(x,y) (x^2+y^2) dxdy}
\end{equation}
where $I$ is the intensity, $W$ is our Gaussian weight, $x$ and $y$ are 
measured from the (weighted) centroid of the image, and the boldface indicates 
that $\bfe$ is a complex quantity.

We correct for the effects of the point spread function (PSF) in two steps. 
First, we correct for the effect of the shape of the PSF by
reconvolving the observed images with a 
spatially varying convolution kernel which is designed to make the 
stars round.  The galaxies in the convolved images are thus no longer
affected by the {\em shape} of the point spread function (PSF), but are 
still affected by the {\em size} of the PSF. 

Since the convolution kernel is only measured where we have an observed
star, and the PSF is far from uniform across each image, 
we need to interpolate between the stars.  In our previous analysis,
we used a separate interpolation for every image.  
For this analysis, however,
we use our new principal component algorithm which uses the information 
from all of the images at once.
This new method, which gives a better fit, is described in \citet{Ja05}.

The effect of the size of the PSF is called dilution.  A perfectly round PSF
blurs the images of galaxies, which reduces the observed ellipticity.
For a purely Gaussian PSF and Gaussian galaxies, the measured ellipticities 
are reduced by a factor $R$:
\begin{equation}
R = 1 - \frac{\sigma^2_{\rm psf}}{\sigma^2_{\rm gal}}
\label{gaussian_response}
\end{equation}

However, galaxies are certainly not Gaussian, and stars are only approximately
Gaussian.  In our previous analysis we used a formula intended to account
fot the first order corrections due to the kurtosis of the galaxies and the
PSF.  Unfortunately, our formula was incorrect with respect to the
PSF kurtosis, as pointed out by \citet{Hi03}.
They give a more accurate correction scheme which accounts for 
both kurtoses correctly to first order -- we implement this scheme
in the analysis presented below. 
Their formula is quite complicated to write, but is
easy to implement, so we defer to their Appendix~B for the relevant
equations. 

Finally, we estimate the shear, $\bfg$, from an ensemble of ellipticities
using the formula:
\begin{equation}
\hat{\bfg} = \frac{1}{2\RR} \frac{\sum w_i \,\bfe_i}{\sum w_i}
\end{equation}
where the responsivity, $\RR$, describes how our mean ellipticity
changes in the presence of an applied distortion. It generalizes 
the formula given in equation \ref{gaussian_response} to 
non-Gaussian shapes. The factor of 2 in the denominator above converts
the ellipticity to shear, and $w$ is our weight function.  We use the ``easy''
weight function given in \citet{BJ02} (Equation 5.36):
\begin{equation}
w = [e^2 + (1.5\sigma_\eta)^2]^{-1/2}
\end{equation}
where $\sigma_\eta$ is the shape uncertainty in the sheared coordinates 
where the galaxy is circular.  The corresponding responsivity, $\RR$, 
is also given in \citet{BJ02} (Equation 5.33).

\section{Weak Lensing Statistics}
\label{statssection}

To describe the two-point statistics of our shear field, we use 
the aperture mass statistic \citep{Sch98,Cr02,Sch02,Pen02}.  

The predictions from theory come in the form of the convergence
power spectrum:
\begin{equation}
P_\kappa(\ell)
= \frac{9}{4} \Omega_0^2 \int_0^{\chi_H} d\chi
  \frac{g^2(\chi)}{a^2(\chi)}
  P_{3D} \left( \frac{\ell}{r(\chi)}; \chi \right)
\label{Pkappa}
\end{equation}
where $\chi$ is the radial comoving distance,
$\chi_H$ is the horizon distance,
$r(\chi)$ is the comoving angular distance,
$a$ is the scale size of the universe,
$P_{3D}$ is the three-dimensional power spectrum of the matter fluctuations,
and
\begin{equation}
g(\chi) = \int_\chi^{\chi_H} d\chi^\prime p(\chi^\prime) 
\frac{r(\chi^\prime-\chi)}{r(\chi^\prime)}
\label{gchi}
\end{equation}
where $p$ is the normalized (to give unit integral over $\chi$)
redshift distribution of source galaxies.
These predictions are not reliable for high values of
$k$ ($k > 10 h{\rm Mpc}^{-1}$) \citep{Sm03} due to difficulties
in predicting the non-linear growth.

The observed second moments are completely
described using the two-point correlation functions: 
\begin{eqnarray}
\xi_+(\theta)
&=& \langle \gamma({\bf r}) \gamma^*({\bf r + \bft}) \rangle \nonumber\\
\xi_-(\theta) + i\xi_\times(\theta)
&=& \langle\gamma({\bf r}) \gamma({\bf r +\bft}) e^{-4i\alpha}\rangle
\end{eqnarray}
where $\bft = \theta e^{i\alpha}$ is the separation between pairs of
galaxies, and treating the positions on the sky as complex values.
The correlation functions are not measured at all on
scales larger than the size of the survey fields (for this survey, at
$\theta > 200\arcmin$), which correspond to low $k$ values. 

Using the
information from the correlation function to obtain the power
spectrum, or vice versa, requires an
extrapolation away from the $k$ values which are well measured or well 
predicted.
The aperture mass statistic is in some sense a compromise between these
two statistics, since it can be calculated from either the power spectrum
or the correlation function using only the range of $k$ values which are
either predicted or measured, respectively:
\begin{eqnarray}
\langle M_{\rm ap}^2 \rangle (R)
\label{mapsq_p}
&=& \frac{1}{2\pi} \int \ell d\ell P_\kappa(\ell) W(\ell R)^2 \nonumber\\
\label{mapsq_xi}
&=& \frac{1}{2} \int \frac{\theta \,d\theta}{R^2}
   \left[ \xi_+(\theta) T_+\left(\frac{\theta}{R}\right)
        + \xi_-(\theta) T_-\left(\frac{\theta}{R}\right)
   \right] 
\end{eqnarray}
where we use the form suggested by \citet{Cr02} for which we have:
\begin{eqnarray}
W(\eta) &=& \frac{\eta^4}{4} e^{-\eta^2} \nonumber\\
T_+(x) &=& \frac{x^4-16x^2+32}{128} e^{-x^2/4} \nonumber\\
T_-(x) &=& \frac{x^4}{128} e^{-x^2/4}
\end{eqnarray}
The aperture mass therefore has both good 
predictions from theory and accurate measurements from the data.

While the integral in \eqq{mapsq_xi} is technically from 0 to infinity,
the $T$ functions drop off very quickly, so that the effective upper
limit is really around $5R$.  Our fields are $200\arcmin$ along the long
diagonal, so we can measure the aperture mass up to $R = 40\arcmin$.  The 
lower limit, due to difficulties of measuring the correlation function
on very small scales, is about $R = 1\arcmin$.

The function $W(\ell R)$ is a narrow function around
$\ell R = \sqrt{2}$, so this corresponds to a range in $\ell$ of approximately
$120 < \ell < 5000$.

The other big advantage to the aperture mass statistic is that it 
provides a natural check for systematics.  Weak lensing 
should only produce a curl-free $E$-mode component, so any $B$-mode observed
in the shear field represents a systematic error of some sort.  
The aperture mass statistic has a $B$-mode counterpart which can 
be likewise calculated from the correlation functions as:
\begin{equation}
\langle M_\times^2 \rangle (R)
= \frac{1}{2} \int \frac{\theta \,d\theta}{R^2}
   \left[ \xi_+(\theta) T_+\left(\frac{\theta}{R}\right)
        - \xi_-(\theta) T_-\left(\frac{\theta}{R}\right)
   \right] 
\end{equation}

\begin{figure}[t]
\epsscale{1.0}
\plottwo{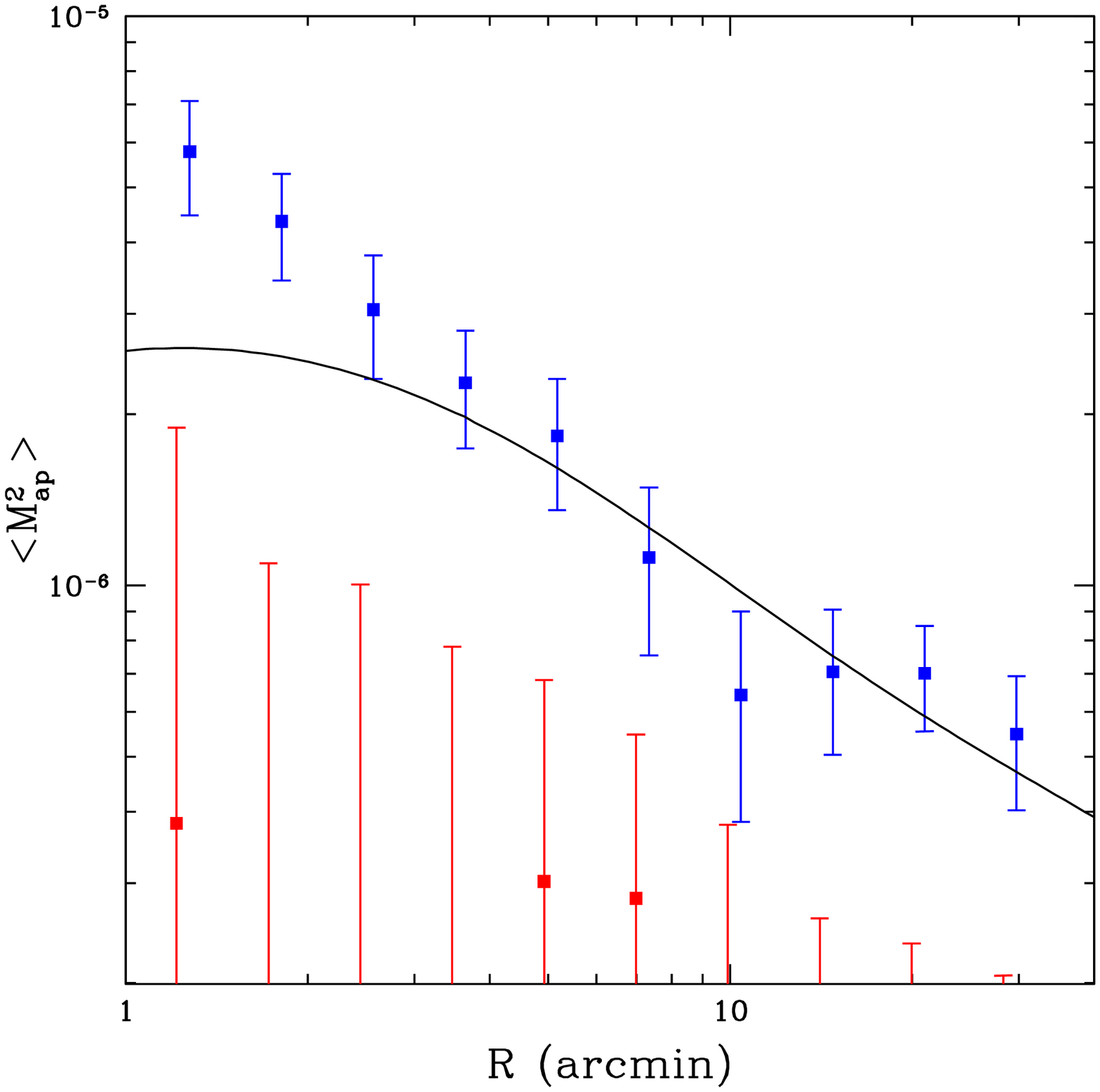}{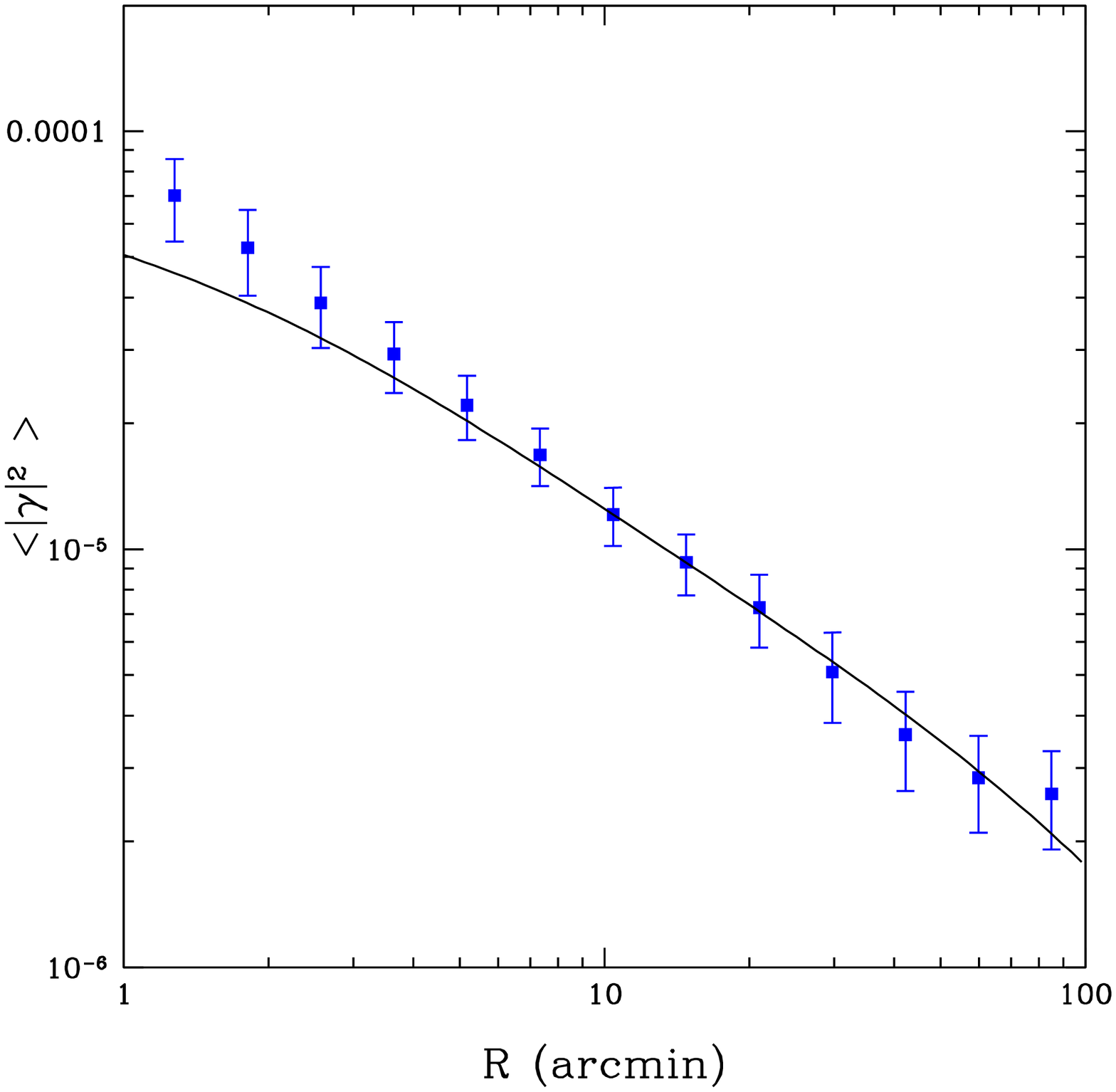}
\caption[]{ \small
The aperture mass (left) and shear variance (right) measurements for our 
CTIO survey data.
For the left plot, the blue (upper) points are the $E$-mode measurements,
$\langle M_{\rm ap}^2 \rangle(R)$ and the red (lower) points are the 
$B$-mode contamination, $\langle M_\times^2 \rangle(R)$.
For the right plot, the blue points are the total shear variance,
including any potential $B$-mode contamination.
In both cases, the black curve is the best fit flat $\Lambda$CDM model found 
in \S\ref{analysissection}.  
}
\label{mapvarfigure}
\end{figure}

Figure~\ref{mapvarfigure} shows the results for 
our reanalysis.  
The blue points are the $E$-mode signal, and 
the red points are the $B$-mode contamination.  Points separated by at least 
one other point are very nearly uncorrelated.  
The black curve is the best fit flat $\Lambda$CDM model found below.

The $B$-mode is seen to be consistent with zero at all scales,
which was not the case for our previous analysis.  Further, in 
Jarvis \& Jain (2004) we show that the measured ellipticity
correlation function of stars, which is another measure of systematic
errors, is one to two orders of magnitude
smaller than the expected lensing signal at all scales. Therefore, we can
now confidently use all of the aperture mass values from $1\arcmin$
to $40\arcmin$ for our constraints on cosmology.

In addition to the aperture mass, we also measure the variance of the mean
shear in circular apertures:
\begin{eqnarray}
\langle|\gamma|\rangle^2(R)
\label{var_p}
&=& \frac{1}{2\pi} \int \ell d\ell P_\kappa(\ell) 
   \frac{4 J_1(\ell R)^2}{(\ell R)^2} \nonumber\\
\label{var_xi}
&=& \int_0^{2R} \frac{\theta \,d\theta}{R^2}
   \xi_+(\theta) S_+\left(\frac{\theta}{R}\right)
\end{eqnarray}
where
\begin{equation}
S_+(x) = \frac{1}{\pi} \left(4 \arccos(x/2) - x \sqrt{4-x^2}\right)
\end{equation}

Figure~\ref{mapvarfigure} shows the results for the shear variance statistic.
There is no $E/B$ decomposition for this statistic\footnote{
Technically, there is \citep{Sch02}, 
but it requires extrapolation of the correlation 
functions past where they are measured.
}. 
We assume implicitly that there is no $B$-mode contamination
in the shear variance measurements, which seems reasonable given the low
$B$-mode seen for the aperture mass.

The usefulness of this statistic is that it is able to probe the power spectrum
at somewhat smaller $\ell$ values than the aperture mass statistic.  The upper 
limit in the integral in \eqq{var_xi} is only $2R$, so we can calculate the shear
variance up to $R = 100\arcmin$ with our data.  This probes the power spectrum
down to $\ell$ of about 70.  This leads to a total dynamic range
for both statistics of almost 2 orders of magnitude.  The shear variance
below $R = 50\arcmin$ is degenerate with the aperture mass,
so for our constraints, we only use the shear variance at large values
of $R$ where it is providing extra information.

In Figure~\ref{mapvarfigure}, we also show the overall best fit 
$\Lambda$CDM model (see \S\ref{analysissection}).  This fit has a 
$\chi^2$ of 7.7, for effectively 6 degrees of freedom, yielding
a reduced $\chi^2$ of 1.28.

\section{Analysis}
\label{analysissection}

\subsection{Dark Energy Constraints from Weak Lensing}

Our lensing measurement constrains the shear power spectrum, 
which is a weighted projection of the mass power spectrum.
The constraints on dark energy arise 
from two sources. The first is the angular diameter distances
to the lens, to the source, and between the lens and the source that enter 
into the projection. The second is the amplitude of the power spectrum. 
The dark energy component alters the expansion rate of the universe
at redshifts below about 2 (at least if its evolution is not too 
different from a cosmological constant). This in turns affects 
the growth of structure. Since the CMB fixes the amplitude of the
power spectrum at $z=1100$, the lensing measurement of the 
amplitude at low redshift measures the growth function.  See
\citet{Hu04} for a detailed discussion.

We assume that massive neutrinos make a negligible contribution 
to the matter density, that the primordial power spectrum index has no running
and that the universe is spatially flat.
The shape of the mass power spectrum is then specified by the baryon 
density $\Omb h^{2}$, the matter density $\Omm h^{2}$,
and the primordial spectrum.
Following the WMAP convention, we use the scalar amplitude
$\As$ and spectral index $n$ such that the shape of the primordial
power spectrum is $\As (k/k_0)^{(n-1)}$,
where $k_0=0.05$ Mpc$^{-1}$ is the normalization scale.  
The current uncertainties in these parameters are at the 10\% level or
better.  Thus the power spectrum as a function of $k$ in Mpc$^{-1}$
(not $h$ Mpc$^{-1}$) in the matter dominated regime can be
considered as largely known.

The amplitude of the power spectrum at a given redshift 
depends on the initial normalization $\As$
and the ``growth function'' $G$ defined by
\begin{equation}
P(k,z) = \left[ \frac{1}{1+z} \frac{G(z)}{G_0} \right]^2 P(k,0) \,,
\end{equation}
where $G_0\equiv G(z=0)$ and
we assume that all relevant scales are sufficiently below the 
maximal sound horizon of the dark energy.  

The normalization of the linear power spectrum today is conventionally given
at a scale of $r=8 h^{-1}$Mpc and can be approximated as \citep{Hu04}
\begin{equation}
\sigma_{8} ~
\approx~ \frac{\As^{1/2}}{0.97}
\left( \frac{ \Omb h^{2}}{0.024} \right)^{-0.33}
\left( \frac{ \Omm h^{2} }{ 0.14} \right)^{0.563} 
~\times(3.123h)^{(n-1)/2} \left( \frac{ h }{ 0.72} \right)^{0.693}  
\frac{G_0 }{ 0.76}\,,
\label{eqn:sigma8}
\end{equation}
Thus a measurement of $\sigma_8$, in conjunction with constraints on
the other parameters in the above equation from the CMB, 
constrains a combination of dark energy parameters that affect $G_0$.
Note that while the equation above illustrates how the dark energy
parameters are linked to others, we do not actually use it in our
parameter analysis. Instead we use the projection integral of
\eqq{Pkappa} which includes the full range of redshifts probed by 
our survey. The peak contribution to the lensing correlations 
is from $z\simeq 0.3$, though the maximum sensitivity to $w$ is
at $z\simeq 0.4$ as discussed below. Combining our measurement 
with others such as Type Ia supernovae, which are sensitive to
different redshifts, enables a probe of the time
dependence of dark energy parameters. With photometric redshifts, 
this could be done with the lensing data alone using the auto 
and cross-spectra in redshift bins. 

The dark energy modifies the expansion of the universe according to 
the equation (for a flat universe)\citep[\eg][]{Li02}:
\begin{equation}
H^2(a) = H_0^2 \left[ \Omm a^{-3} + \Omde e^{
    -3 \int_1^a {\frac{da}{a} (1 + w(a))} } \right]
\end{equation}
where the dark energy density is 
$\Omde(a)=8\pi G \rho_{\rm de}/3H(a)^2$,
its equation of state is 
$w(a)=p_{\rm de}/\rho_{\rm de}$, 
and we indicate the present time values, $\Omm(a=1)$ and $\Omde(a=1)$
as simply \Omm and \Omde.

Distances are then given by 
\begin{equation}
\chi(a) = \int_a^1 { \frac{da^\prime c}{H(a^\prime){a^\prime}^2} }
\end{equation}
and the growth function $G$ depends only on the dark energy
throught the equation:
\begin{equation}
\frac{d^2 G}{d \ln a^2} + 
\left[ \frac{5}{2} - \frac{3}{2} w(a) \Omde(a) \right]
\frac{d G}{d \ln a}   
+ 
\frac{3}{2}[1-w(a)]\Omde(a) G =0\, .
\end{equation}

When we use a constant $w$ parameterization, it is equivalent to 
a measurement of $w$ at the redshift for which the errors in 
the constant and time-dependent piece (in a Taylor expansion)
are uncorrelated. See \citet{Hu04} for a discussion of 
this pivot redshift, which we find (\S\ref{varwsection})
to be about $0.4$ for our survey
(when combined with the CMB and SN priors as described below).

\subsection{Joint Constraints}

We use the results from WMAP \citep{Sp03}
as priors for our analysis.  Specifically, we use
the Monte Carlo Markov Chain (MCMC) calculated by 
\citet{Ve03}\footnote{
Available at http://www.physics.upenn.edu/~lverde/MAPCHAINS/mcmc.html.
}.
We choose to use the less restrictive prior of $w > -3$ rather than 
$w > -1$ in order to be as conservative as possible. However, we do have
a hard prior that $\Omk = 0$.
While WMAP has constrained this to be
$0.02 \pm 0.02$ for pure $\Lambda$CDM, and there is theoretical bias in 
believing it is
exactly 0, it is worth emphasising that our dark energy 
constraints would be weakened if this prior were relaxed.

Each step in the Markov chain contains a value for each of the following
parameters:
$\omega_{\rm b} = \Omb h^2$ is the density of baryons;
$\omega_{\rm c} = \Omc h^2$ is the density of cold dark matter;
$\theta_{\rm A}$ is the angular scale of the acoustic peaks;
$n$ is the spectral slope of the scalar primordial density power spectrum;
$Z = \exp(-2\tau)$ is related to the optical depth ($\tau$) of the last 
scattering surface;
$\As$ is the overal amplitude of the scalar primordial power spectrum;
$h = H_0/100$ is the Hubble constant;
$w$ is the equation of state parameter for the dark energy.

The parameters $\theta_A$ and $Z$ are not directly 
relevant to the aperture mass 
statistic which we have measured, so we marginalize over these two.  
The others define a cosmology from which we can predict the aperture 
mass statistic on the scales where we have measured it. We present
results for selected parameters below, these are obtained by a full
marginalization over all the other parameters listed above. 

For the linear power spectrum, we use the transfer function of 
\citet{BBKS}.
We then estimate the non-linear power spectrum using the 
halo-based model of model of \citet{Sm03}.  
Their fitting formulae provide a means of estimating the quasi-linear
and non-linear halo contributions to the power spectrum based on
the linear value and the effective spectral index.  This model 
agrees with the results of $N$-body simulations better than the
simpler formula of \citet{PD96}.
The nonlinear correction affects the predicted variance in the mass
aperture statistic on scales below about 4 arcminutes.

We also note that the model of \citet{Sm03} that we use does not include
$w$ directly.  We correctly take it into account for the growth factor
and the distances, but \citet{Ma03} and \citet{Kl03} show that 
dark energy changes the virial density contrast, $\Delta_c$, which
results in changes in the power spectrum at high
$k$ values ($k > 1\ h {\rm Mpc}^{-1}$).  However, the effect is smaller than
the expected error in the non-linear model even for a $w \simeq -0.5$ 
cosmology.

From the predicted power spectrum, we calculate the 
aperture mass and shear variance statistics using 
Equations~\ref{Pkappa}, \ref{mapsq_p}, and \ref{var_p}.
Our data then give a likelihood value for each cosmology, which we
combine with the CMB likelihood from the MCMC. 

We also use the recent supernova measurements of \citet{Ri04} to further
constrain the results.  For this, we use the $\chi^2$ calculation program of
\citet{To03}\footnote{
Available at http://www.ifa.hawaii.edu/~jt/SOFT/snchi.html.
The code there was modified slightly to read the data of \citet{Ri04}.
}.

\subsubsection{$\Lambda$CDM Models}

\begin{figure}[t]
\epsscale{1.0}
\plottwo{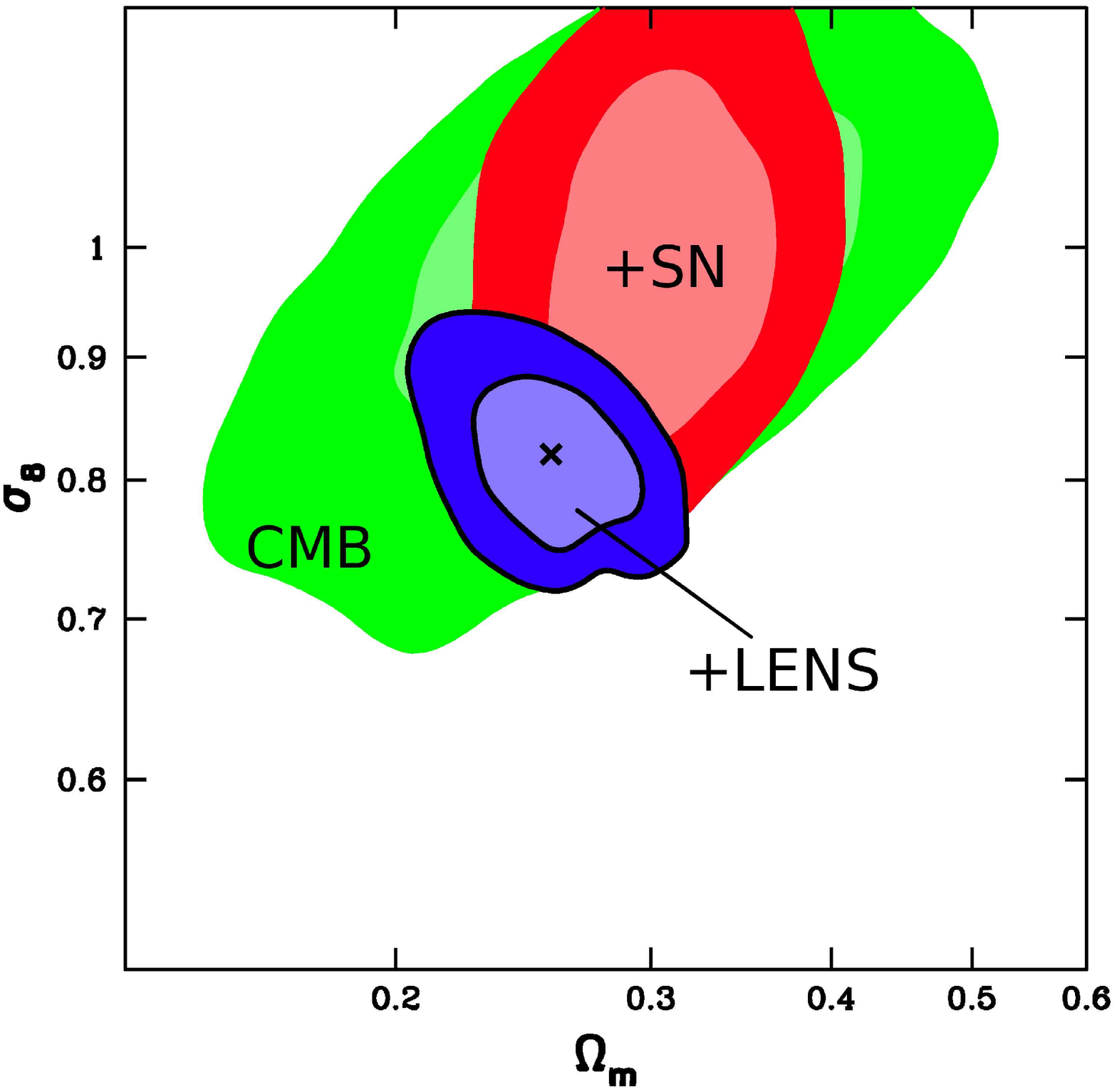}{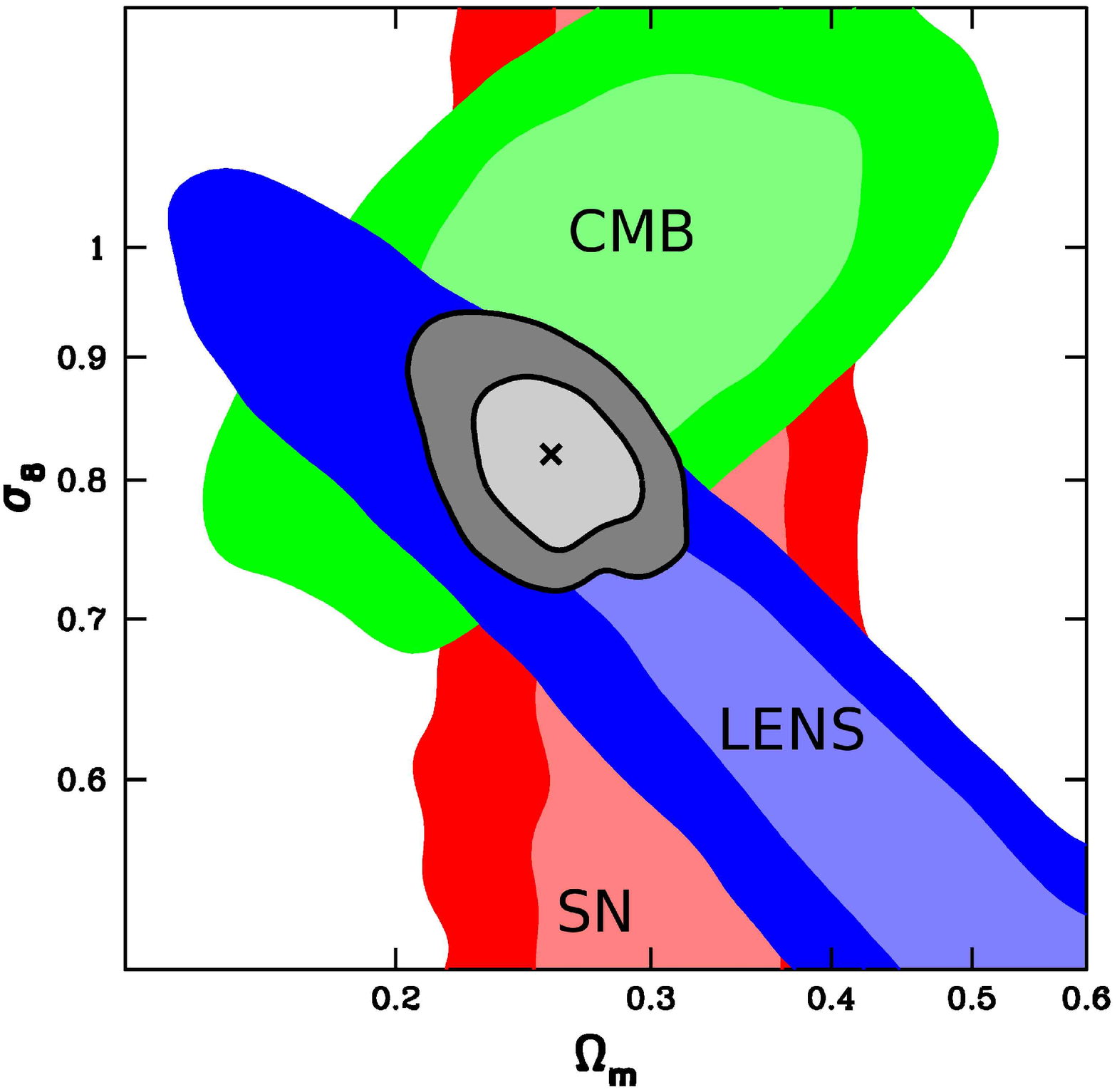}
\caption[]{ \small
Contour plots of $\chi^2$ for the $(\Omm,\sigma_8)$ plane.
The left plot shows the effect of adding the data sets sequentially,
starting with the CMB constraints, then adding the supernova and 
lensing data.
The right plot shows contours for each of the three data sets separately.
In each case the contours
enclose the 68\% and 95\% confidence regions.  The $\times$ is the
best fit model. All other parameters are marginalized over as
discussed in the text. 
}
\label{lambda_plots}
\end{figure}

With dark energy priors of $w = -1$ and $w_a = 0$,
the likelihood constraints are non-trivial contours through a 
five-dimensional parameter space.  However, most of the gain in constraining
power from the addition of the lensing data comes in the quantities $\Omm$ 
and $\sigma_8$.
We show the error contours projected onto the $\Omm-\sigma_8$
plane in Figure~\ref{lambda_plots}.
The left plot shows the contours starting with
the CMB data set, and sequentially adding the supernova and lensing data.
The right plot shows the contours in the $\Omm - \sigma_8$
plane separately for each of the three data sets to indicate the degree
of their complementarity, which is why their combination
leads to the tight overall constraints.
In both plots, the contours correspond to 68\% and 95\% confidence regions
($\Delta \chi^2$ = 2.30 and 6.17).  The crosses are at the peak
likelihood in the projected plane, which is at: 
($\Omm = 0.26$, $\sigma_8 = 0.82$).

\subsubsection{Constant $w$ Models}

\begin{figure}[t]
\epsscale{1.0}
\plottwo{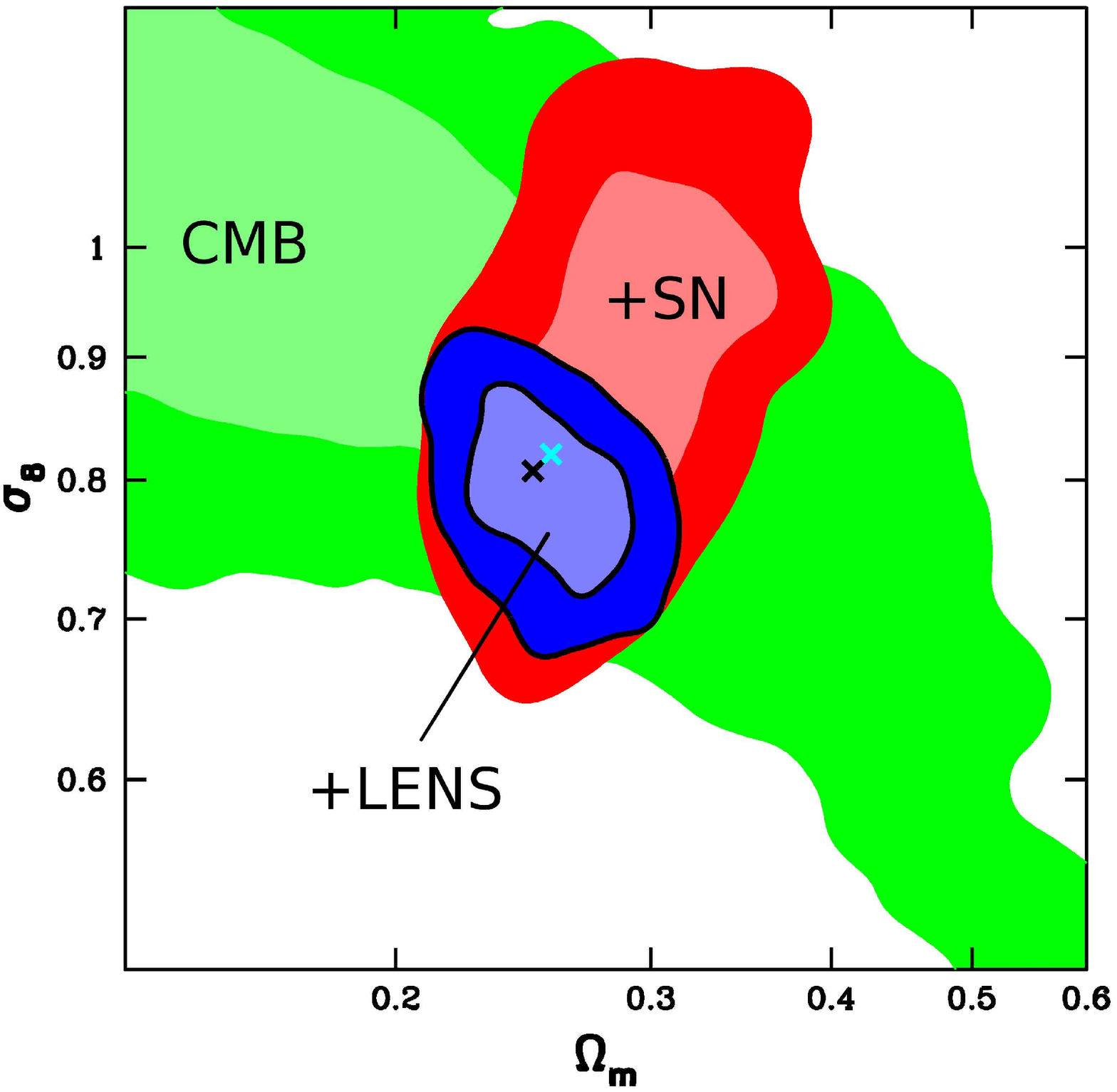}{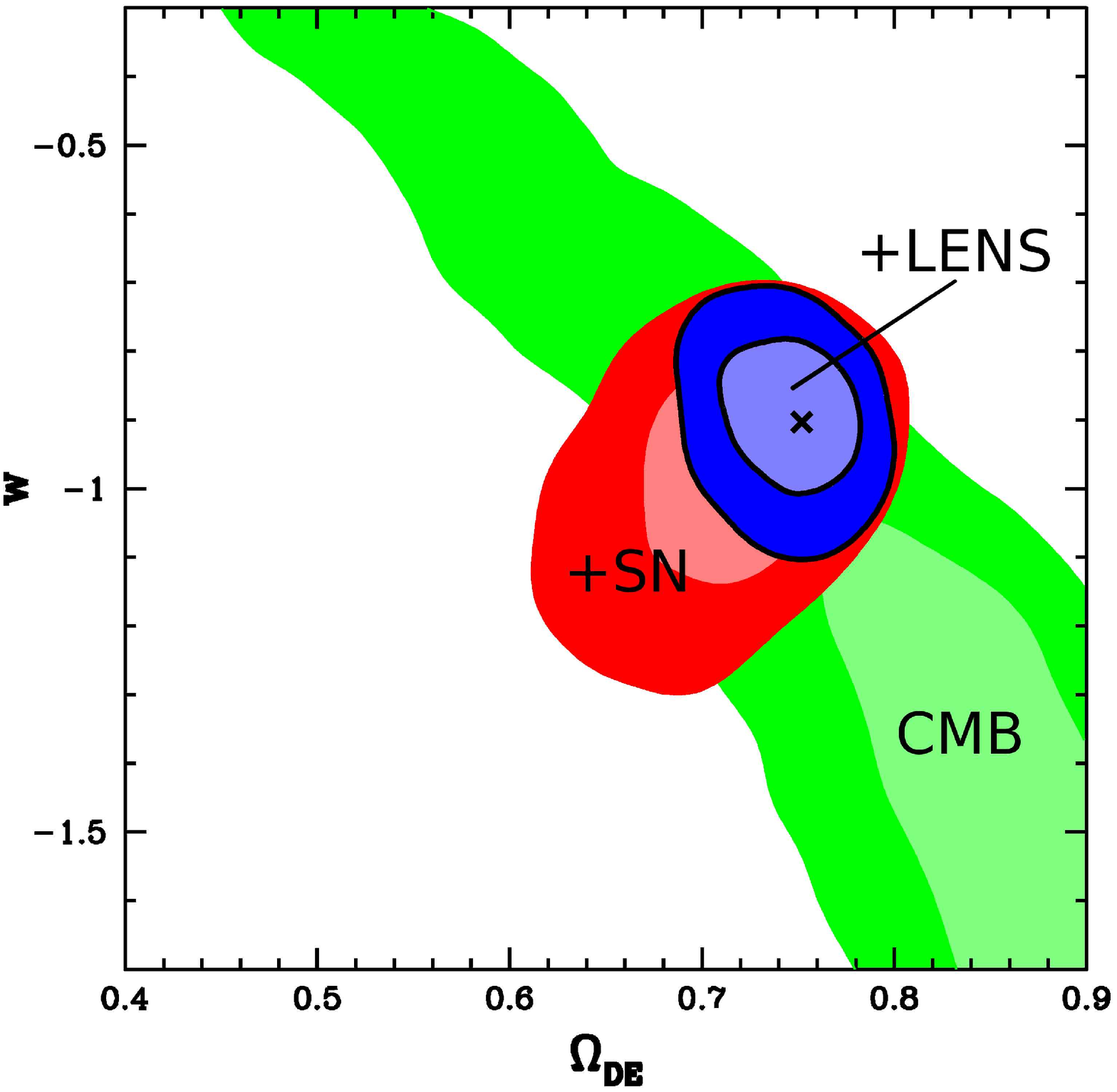}
\caption[]{ \small
Contour plots of $\chi^2$ for the $(\Omm,\sigma_8)$ plane (left)
and the $(\Omde,w)$ plane (right) for the constant $w$
dark energy models.
Both plots show the effect of adding the data sets sequentially.
In each case the contours
enclose the 68\% and 95\% confidence regions.  The black $\times$'s are the
best fit models in each plane.  The cyan $\times$ in the left plot is
the best fit from the $\Lambda$CDM prior (Figure~\ref{lambda_plots}).
}
\label{w3_plots}
\end{figure}

We show the uncertainty contours for the dark energy priors of  
$-3 < w_0 < 0$ and $w_a = 0$ in Figure~\ref{w3_plots}.
The left plot give the projection in the $\Omm-\sigma_8$ plane,
which indicates that allowing $w$ to be free does not 
significantly worsen the constraints on $\Omm$ and $\sigma_8$ compared
to the pure $\Lambda$CDM model. 
The $\Omde-w$ plot (right) shows why.  While none of the three data sets 
individually have tight constraints in this plane, the combination of all
three leads to a fairly tight contour near (and consistent with) $w=-1$.
The peak likelihood models in the two projections are:
($\Omm = 0.25$, $\sigma_8 = 0.79$) and 
($\Omde = 0.75$, $w = -0.90$).

\subsubsection{Variable $w$ Models}
\label{varwsection}

\begin{figure}[t]
\epsscale{1.0}
\plottwo{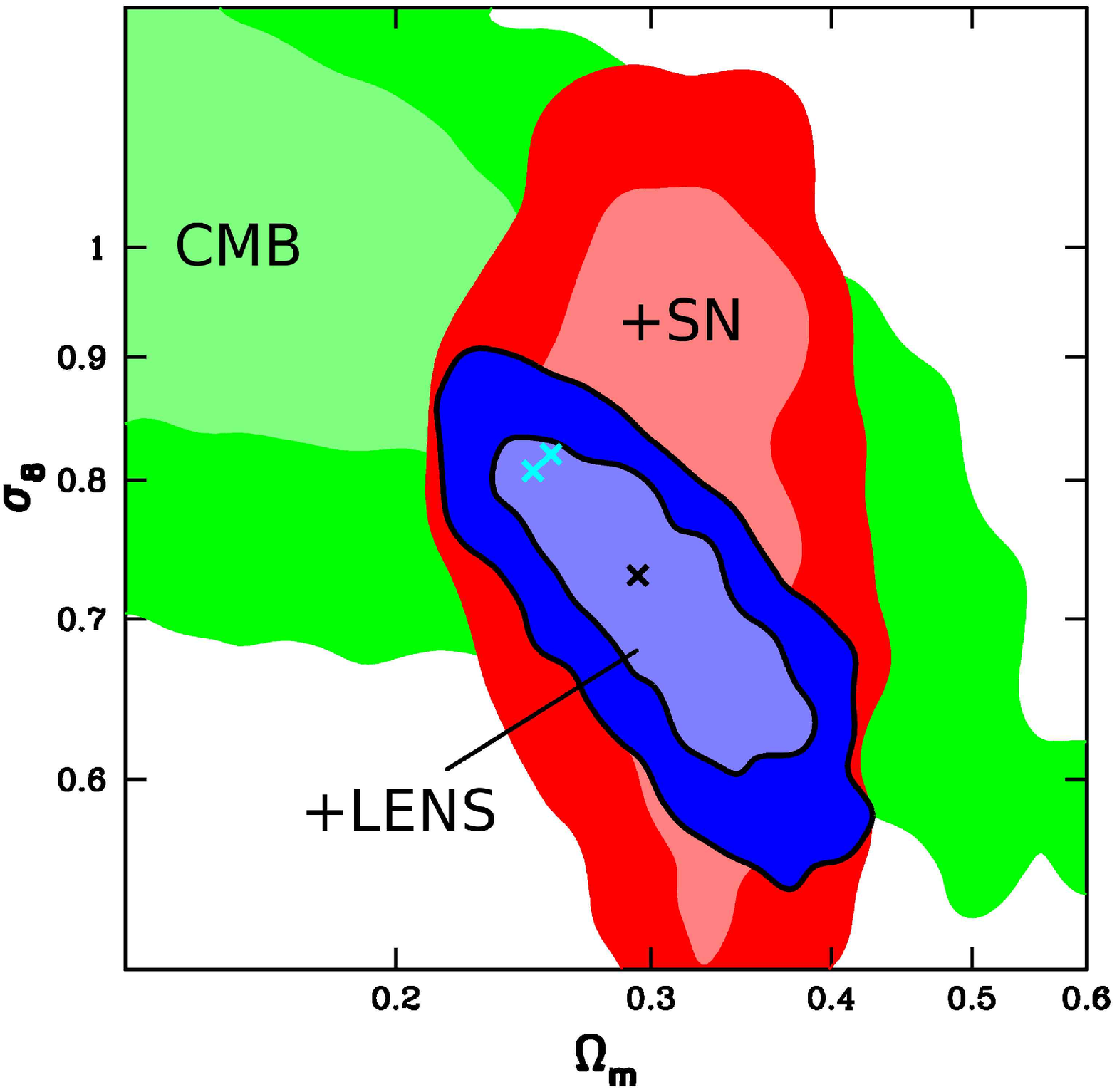}{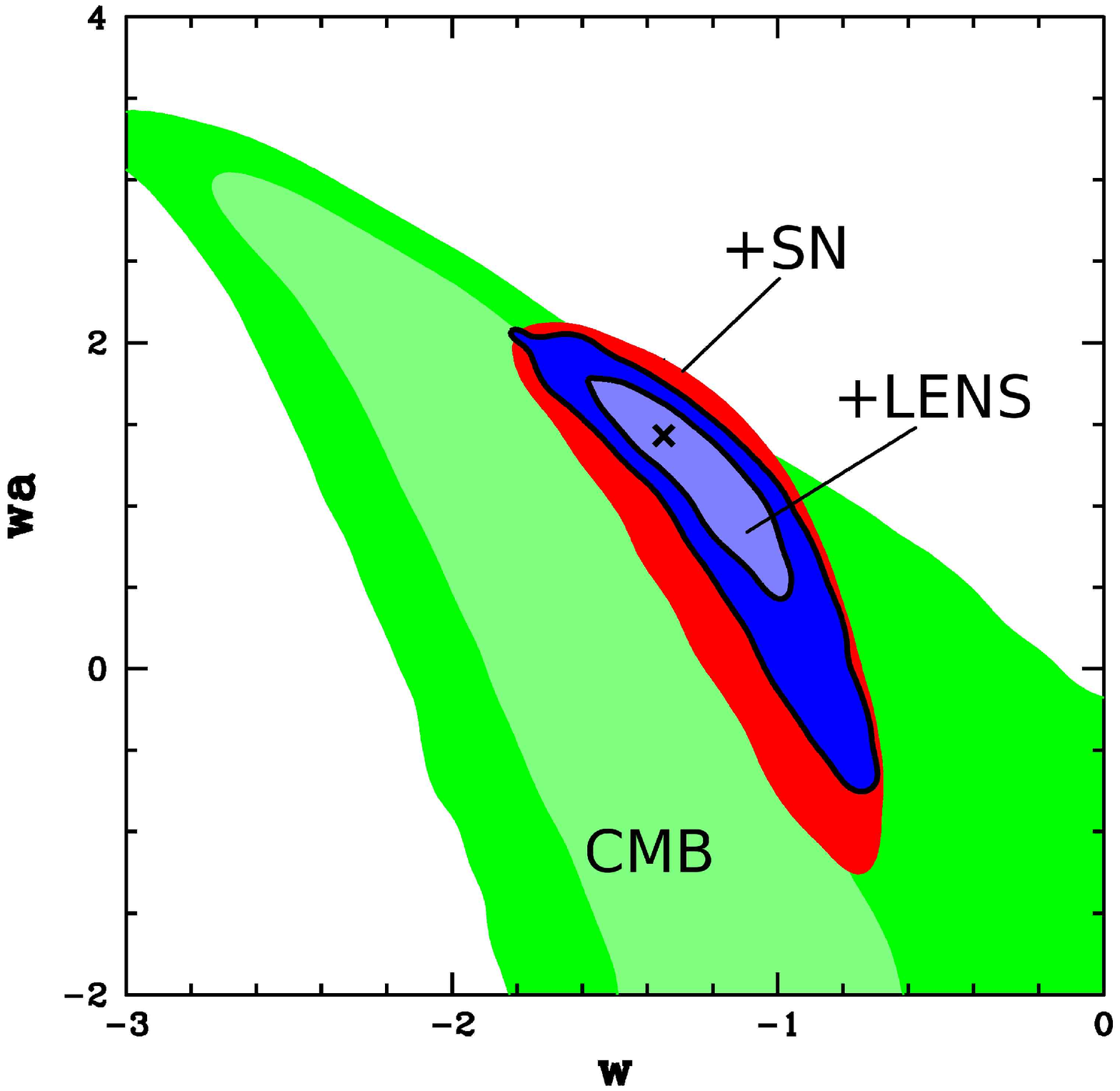}
\caption[]{ \small
Contour plots of $\chi^2$ for the $(\Omm,\sigma_8)$ plane (left)
and the $(w_0,w_a)$ plane (right) for the variable $w$
dark energy models.
Both plots show the effect of adding the data sets sequentially.
In each case the contours
enclose the 68\% and 95\% confidence regions.  The black $\times$'s are the
best fit models in each plane.  The cyan $\times$'s in the left plot are
the best fits from the $\Lambda$CDM and constant $w$ priors 
(Figures~\ref{lambda_plots} and \ref{w3_plots}).
}
\label{wwa_plots}
\end{figure}

Finally, we consider dark energy priors of $-8 < w_0 < 8$ and $-8 < w_a < 8$.
It turns out that some of the dark energy models in this range have 
$\Omega_{\rm DE}(z=1100) \approx 1$.  That is, the mass-energy of the 
universe was essentially all dark energy at the epoch of recombination.
This seems to be ruled out by WMAP data \citep{Ca03,Ca04,Wa04}.
Therefore, we make the additional prior that $\Omega_{\rm DE}(z=1100) < 0.5$.
In practice, all the models have $\Omega_{\rm DE}(z=1100) \approx 0$ or $1$, 
so this contraint is effectively $\Omega_{\rm DE}(z=1100) \approx 0$.

Given this constraint, the primary effect of dark energy on the CMB
is through the distance to the last-scattering surface, $d_{\rm LSS}$.
Therefore, we approximate the CMB likelihoods by using the WMap
constant-$w$ Markov chain mentioned
above, modifying the dark energy parameters to maintain a constant 
$d_{\rm LSS}$.  Specifically, for each line in the Markov chain,
we determine $d_{\rm LSS}$ from the values of $\Omega_m$ and $w$;
we select $w_a$ from $-8 < w_a < 8$; then we determine 
what $w_0$ with this $w_a$ and the same $\Omega_m$ maintain the 
given value of $d_{\rm LSS}$,
and we write these values out as a line in a new pseudo-chain.

The main approximation in this process is that we neglect the difference
of the integrated Sachs-Wolfe (ISW) effect between the two models. 
There is some indication\footnote{
Wayne Hu, private communication
} that for some of the models that fall within our contours 
(Figure~\ref{wwa_plots}, right),
the ISW may spread the first peak in the power spectrum of the
CMB enough to disfavor these models by 1 or 2 sigma.  However, we defer 
a more detailed study of this effect to future work.

The error contours projected onto the $\Omm-\sigma_8$
plane and the $w_0-w_a$ plane are shown in Figure~\ref{wwa_plots}.
With this much freedom in the models, 
we do finally lose much of our constraining power.  Even with all 
three data sets, the error contours are still quite large.
The $\Omm-\sigma_8$ plot (left) shows that 
allowing $w$ to vary with cosmological time does
significantly worsen the constraints, allowing much lower values of 
$\sigma_8$.

Also, while lensing does significantly reduce the allowed parameter
space in the $\Omm-\sigma_8$ plane, this reduction has only a small
effect in the $w_0-w_a$ plane, so that our constraints there are 
not much better than those for the CMB + SN data alone.
The data are consistent with a cosmological constant
($w_0=-1, w_a=0$) at the 95\% confidence level, 
but $w_a$ may range as high as 2\footnote{
It is for some of these high $w_a$ models that the ISW effect may be
important, requiring a more careful analysis to determine what 
portion of the nominally allowed region is really ruled out.}.

We can use the direction of the contours in the $w_0-w_a$ plane to determine
the redshift at which we have the strongest constraints on the
dark energy.  If we change variables from $\{w(z=0),w_a\}$ to 
$\{w(z=0.4),w_a\}$,
the likelihood contour becomes roughly vertical.  
This indicates that
our pivot redshift, or ``sweet spot'' \citep{Hut01,We02,Hu02,Hu04}, 
where the constraints of the dark energy are strongest,
is at a redshift of about $0.4$.
(Of course, the 
banana shape makes this impossible to do precisely, 
so $z_{\rm piv}=0.4$ is just
an approximate value.)  
We constrain $w$ at this redshift, 
marginalizing over $w_a$ (and everything else) below.

\subsection{Systematic Errors}

Systematic errors are harder to estimate than statistical errors, since
by their nature they contaminate the data in unknown ways.  There are
four systematic errors which we investigate and attempt to estimate: 
residual anisotropic PSF as estimated by the residual $B$-mode,
calibration error, redshift distribution error,
and errors in the non-linear prediction.  For shorthand, we refer
to these as B, CAL, Z, and NL respectively.

When estimating the contributions of these systematic uncertainties to 
our error budget,
we limit our consideration to constraints on single parameters, 
fully marginalized over all the other parameters.
First we calculate the 95\% error bars with only the statistical errors.
Then, for each systematic effect, we change our handling of the effect
as described more completely below for each case.  When we do this,
the 95\% confidence intervals move around somewhat.  We define the
upper systematic error to be the maximum upper limit of the confidence
interval allowed by the various changes
minus the nominal upper limit with only the statistical errors.
Likewise the lower systematic error is the lower limit with only the 
statistical errors minus the minimum allowed lower limit.  Finally, we
(conservatively) estimate the total error as the sum of the statistical
errors and each of the systematic errors added linearly, not in quadrature.

For the residual PSF (B) systematic, we implement the same technique
we used in \citet{Ja03}, namely running the analysis with the $B$-mode
contamination added to and subtracted from the $E$-mode signal.
Most types of contamination either add power (roughly) equally to
the $E$ and $B$ modes, in which case the subtraction is appropriate,
or mix power between the two modes while conserving total signal, 
in which case the addition is appropriate.  We allow for both 
possibilities to estimate how the contamination could be affecting
the cosmological fits.  

The calibration (CAL) uncertainty includes errors in the dilution calculation,
the responsivity formula, and possibly biases in the shape measurements.
This systematic was the subject of significant discussion at the recent
IAU symposium 225\footnote{July 19-23, 2004, Lausanne, Switzerland}.
There seem to be calibration differences of order 5\% in shear
estimators between different methods.  Tests with simulated images
with known shears indicate that we have calibration errors of less than
2\% \citep{STEP}, but we allow for $\pm$ 5\% in our shear values as a 
conservative estimate of this systematic.

For the redshift calibration (Z) of our survey, there
are two public redshift surveys with depths similar to our observations:
the Caltech Faint Galaxy Redshift Survey \citep{CRS} (CRS),
and the Canada-France Redshift Survey \citep{CFRS} (CFRS).
We argue in \citet{Ja03} that the CRS is a better choice, since it is
more complete in the $R$ filter band pass used for our observations.  
However, switching to the CFRS distribution allows us to estimate the 
uncertainties due to the redshift calibration.  Also, 
since we only have one other survey to use, we cannot run symmetric
plus and minus versions of this test.  
So when the 95\% confidence limit moved inward for a value, 
we take the absolute value of the change as the measure of the systematic 
error, since a different redshift survey {\it might} have moved the limit 
a similar amount outward.

Finally, for the non-linear predictions (NL), we used the \citet{Sm03}
model, which were an improvement over that of \citet{PD96}.
Switching back to the older model should give us a rough
(over-)estimate of the remaining uncertainties due to the
non-linear modelling.  Again, we cannot run symmetric tests, so 
we take the absolute value of any change as a measure of the systematic error.

Since this technique is non-standard and may be confusing, an example 
with the actual values might help explain it.
For the CAL test with the $\Lambda$CDM prior, 
when we multiplied the shear data by 1.05 (for the
+5\%\ test), the upper limit of the 95\%\ confidence interval for $\sigma_8$ 
moved from 0.910 to 0.929, and increase of 0.019.  
This is our estimate of the positive 
systematic error.  Similarly, in the -5\%\ CAL
test, the lower limit decreased, providing the negative systematic error 
for $\sigma_8$. 

In Table~\ref{paramstable},
we present the statistical and systematic error estimates for 
$\Omm$, $\sigma_8$, $w_0$ and $w_a$, for each of our dark energy priors.
The first two columns present the estimates for each value
with and without the lensing data,
marginalized over all other parameters, 
and quoting only the statistical errors.
The next four columns show the estimated systematic error due to each 
effect listed above.  The final column includes the total uncertainty
with the systematic errors
added linearly with the statistical uncertainty.
Note that we make no attempt to estimate the systematic errors present in 
the CMB or SN data.

\begin{deluxetable}{clllllll}
\tablewidth{0pt}
\tablecaption{Fully Marginalized Parameter Constraints (95\%\ c.l.)}
\tablecolumns{8}
\tablehead{
\colhead{Parameter} &
\colhead{CMB+SN} &
\colhead{+Lensing} &
\multicolumn{4}{c}{Estimated Systematic Errors} &
\colhead{Final} \\
\colhead{} &
\colhead{Constraint} &
\colhead{(stat. only)} &
\colhead{B} &
\colhead{CAL} &
\colhead{Z} &
\colhead{NL} &
\colhead{Estimate}
}
\startdata
\sidehead{$w_0=-1$, $w_a=0$ prior:} 
$\Omm$ & 
    \phs$0.301^{+0.07}_{-0.07}$ &      
    \phs$0.256^{+0.05}_{-0.05}$ &      
    ${}^{+0.005}_{-0.001}$ &             
    ${}^{+0.008}_{-0.003}$ &             
    ${}^{+0.008}_{-0.005}$ &             
    ${}^{+0.000}_{-0.000}$ &             
    \phs$0.256^{+0.07}_{-0.06}$ \\     
$\sigma_8$ & 
    \phs$0.953^{+0.27}_{-0.18}$ &      
    \phs$0.812^{+0.10}_{-0.09}$ &      
    ${}^{+0.028}_{-0.006}$ &             
    ${}^{+0.019}_{-0.006}$ &             
    ${}^{+0.020}_{-0.007}$ &             
    ${}^{+0.001}_{-0.001}$ &             
    \phs$0.812^{+0.17}_{-0.11}$ \\     
\sidehead{$-3<w_0<0$, $w_a=0$ prior:} 
$\Omm$ & 
    \phs$0.285^{+0.08}_{-0.07}$ &      
    \phs$0.254^{+0.05}_{-0.04}$ &      
    ${}^{+0.005}_{-0.001}$ &             
    ${}^{+0.006}_{-0.005}$ &             
    ${}^{+0.006}_{-0.001}$ &             
    ${}^{+0.000}_{-0.000}$ &             
    \phs$0.254^{+0.07}_{-0.05}$ \\     
$\sigma_8$ & 
    \phs$0.846^{+0.29}_{-0.17}$ &      
    \phs$0.790^{+0.11}_{-0.10}$ &      
    ${}^{+0.022}_{-0.015}$ &             
    ${}^{+0.018}_{-0.012}$ &             
    ${}^{+0.019}_{-0.012}$ &             
    ${}^{+0.001}_{-0.001}$ &             
    \phs$0.790^{+0.17}_{-0.14}$ \\     
$w$ & 
    $-0.935^{+0.16}_{-0.28}$ &         
    $-0.894^{+0.14}_{-0.16}$ &         
    ${}^{+0.006}_{-0.016}$ &             
    ${}^{+0.005}_{-0.017}$ &             
    ${}^{+0.005}_{-0.019}$ &             
    ${}^{+0.001}_{-0.000}$ &             
    $-0.894^{+0.16}_{-0.21}$ \\        
\sidehead{$-8<w_0<8$, $-8<w_a<8$, $\Omega_{\rm DE}(z=1100)<0.5$ prior:}
$\Omm$ & 
    \phs\phn$0.29^{+0.11}_{-0.06}$ &     
    \phs\phn$0.29^{+0.11}_{-0.07}$ &     
    ${}^{+0.009}_{-0.000}$ &               
    ${}^{+0.001}_{-0.000}$ &               
    ${}^{+0.009}_{-0.001}$ &               
    ${}^{+0.000}_{-0.000}$ &               
    \phs\phn$0.29^{+0.13}_{-0.07}$ \\    
$\sigma_8$ & 
    \phs\phn$0.79^{+0.32}_{-0.35}$ &     
    \phs\phn$0.74^{+0.13}_{-0.17}$ &     
    ${}^{+0.044}_{-0.029}$ &               
    ${}^{+0.013}_{-0.016}$ &               
    ${}^{+0.024}_{-0.000}$ &               
    ${}^{+0.000}_{-0.000}$ &               
    \phs\phn$0.74^{+0.21}_{-0.22}$ \\    
$w_0$ & 
    \phn$-1.17^{+0.50}_{-0.57}$ &        
    \phn$-1.14^{+0.41}_{-0.53}$ &        
    ${}^{+0.008}_{-0.044}$ &               
    ${}^{+0.006}_{-0.020}$ &               
    ${}^{+0.016}_{-0.004}$ &               
    ${}^{+0.002}_{-0.000}$ &               
    \phn$-1.14^{+0.45}_{-0.60}$ \\       
$w_a$ & 
    \phs\phn$1.36^{+0.75}_{-1.79}$ &     
    \phs\phn$1.48^{+0.53}_{-1.57}$ &     
    ${}^{+0.101}_{-0.258}$ &               
    ${}^{+0.021}_{-0.055}$ &               
    ${}^{+0.038}_{-0.124}$ &               
    ${}^{+0.013}_{-0.082}$ &               
    \phs\phn$1.48^{+0.70}_{-2.09}$ \\    
$w(z=0.4)$ & 
    \phn$-0.90^{+0.20}_{-0.33}$ &        
    \phn$-0.87^{+0.13}_{-0.24}$ &        
    ${}^{+0.005}_{-0.047}$ &               
    ${}^{+0.000}_{-0.028}$ &               
    ${}^{+0.002}_{-0.007}$ &               
    ${}^{+0.000}_{-0.001}$ &               
    \phn$-0.87^{+0.14}_{-0.32}$ \\       
\enddata
\label{paramstable}
\end{deluxetable}

It is apparent that the systematic uncertainties in our survey are 
smaller than the statistical uncertainties for each of the above cosmological 
parameters. However, there is still significant room for improvement 
in all of these sources of systematic errors.  Future lensing surveys
which expect to reduce the statistical uncertainties by a factor of order 10
will need to address these systematics so that they do not dominate the
final error budget.  And while the systematic errors due to the
non-linear modelling are essentially completely negligible for our survey, 
they will be more significant for lensing surveys with smaller fields than
ours.

\section{Discussion}
\label{discussionsection}

We have used our measurement of the shear two-point correlations
to constrain the clustering of mass at redshifts $z\sim 0.3$ and
the density and equation of state of dark energy. This has been done
by combining the lensing information with the 
CMB and supernovae data. The three probes
are sufficiently complementary that the joint contraints are
significantly better than from any one or two methods. 

We find that the primary sytematic effects on our lensing data
are the redshift distribution of the galaxies, the
overall calibration of the shear estimates, and
the systematic error due to the coherent PSF anisotropy.  With our
new analysis, the total effect of these three systematics is smaller 
than the statistical errors.
The PCA technique has substantially reduced the contribution of the 
$B$-mode, which used to be the dominant systematic error \citep{Ja03}.
We are continuing to work on methods to reduce this and the calibration
uncertainty.

Improving the redshift distribution would require more data: either a 
larger redshift survey of similar depth as our data, or imaging
the galaxies in
three or four other filters to measure photometric redshifts. As
discussed below, this second option would also allow us to bin the galaxies 
and make tomographic measurements. 

We have necessarily made some choices of priors and datasets 
in our parameter analysis. The datasets we have used in addition
to the lensing data are the WMAP first year extended data \citep{Ve03}
(which also include CBI and ACBAR data) and the \citet{Ri04}
Type 1a Supernovae data. 
We have assumed that the universe is spatially flat; weakening this
assumption significantly weakens constraints on dark energy, especially
if $w$ is allowed to vary in time.  We have
also assumed no tensor contribution to the CMB power spectrum,
that the primordial power spectrum is an exact power law (no running),
and we have neglected the effects of massive neutrinos on the power
spectrum. Current upper limits from cosmology \citep[see e.g.][]{Se05}
are below 1 electron volt for the the sum of neutrino masses. 
Allowing for massive neutrinos could lead to 
(at most) a few percent increase in our estimated $\sigma_8$, 
as the presence of massive neutrinos suppresses the power spectrum
on scales that affect our observed shear correlations, but would not
lead to interesting constraints on the neutrino mass. 

Combining our data with CMB and SN data, our investigation of dark energy
models show no evidence for the dark energy being different from a 
standard cosmological constant.  Constant $w$ models are consistent with
$w=-1$, and variable $w$ models are consistent with $w_a=0$.
The constraints on dark energy from weak lensing come from
a range of redshifts centered at $z\sim 0.3$, and extending
by about $0.2$ in redshift on either side. Thus the measurements
of its density and of $w$ should be interpreted with this redshift
range in mind. When we combine our data with supernova data, we
are using information from different redshifts, and the combined data
have a pivot redshift of about $0.4$, where $w$ is best measured.  At this
redshift, we also find that $w$ is consistent with $-1$.

Our analysis can be compared with other recent work that combines
CMB and supernova data with galaxy clustering, the abundance of 
galaxy clusters, the clustering of the Lyman alpha forest and other
probes \citep{Sp03,We03,Te04,Wa04,Se05,rapetti}. 
It is a powerful consistency check that these
different methods appear to agree in their conclusions. 
It is interesting to compare the different redshift ranges probed by these
methods, and explore constraints on the time dependence of the
equation of state \citep[\eg][]{Li03,Hut05}. 

The prospects for constraining dark energy with future lensing
surveys are very interesting. With a well designed survey and
the recent analysis techniques, it can be hoped that systematic
errors would stay at levels comparable to statistical errors. 
The addition of tomographic information with photometric redshifts
would allow for significantly better constraints on cosmological
parameters from lensing alone \citep{Hu99}. 
The use of three point correlations would allow
for some independent checks on systematics as well as 
improved constraints on cosmological parameters 
(see \citealt{Pen03,Ja04} for detections of the lensing skewness, 
and Takada \& Jain 2004 for forecasts). 
Thus even with a survey of size similar
to ours, significant improvements in parameter measurements are
possible. With future surveys that will cover a significant fraction
of the sky, weak lensing should allow for very precise measurements
of the mass power spectrum, the dark energy density and its evolution. 

\acknowledgements
We thank Wayne Hu, Eric Linder, Masahiro Takada
and Martin White for helpful discussions. 
We are grateful to Licia Verde and the WMAP
team for making available their Markov chains and to Adam Riess, 
John Tonry and the High-z Supernova team for making available their 
data and likelihood code. 
We also thank the anonymous referee for useful comments.
This work is supported in part by NASA grant NAG5-10924, NSF grant
AST-0236702, and a Keck foundation grant.

\end{document}